\documentclass[12pt]{article}
\usepackage{amsmath}
\usepackage{amssymb}
\usepackage{graphicx,psfrag,epsf}
\usepackage{enumerate}
\usepackage{natbib}
\usepackage{url} 
\usepackage[]{algorithm2e}
\usepackage{booktabs}
\usepackage{multirow}
\usepackage{tabularx}
\DeclareMathOperator*{\argmin}{argmin}
\newcommand{\blind}{1}

\begin{document}

\def\spacingset#1{\renewcommand{\baselinestretch}%
{#1}\small\normalsize} \spacingset{1}


\if1\blind
{
  \title{\bf Evaluation of Combinatorial Optimisation Algorithms for c-Optimal Experimental Designs with Correlated Observations}
  \author{Samuel I. Watson\thanks{
    The work was funded by MRC grant MR/V038591/1} and Yi Pan\hspace{.2cm}\\
    Insitute of Applied Health Research,
University of Birmingham, UK}
  \maketitle
} \fi

\if0\blind
{
  \bigskip
  \bigskip
  \bigskip
  \begin{center}
    {\LARGE\bf Title}
\end{center}
  \medskip
} \fi

\bigskip
\begin{abstract}
We show how combinatorial optimisation algorithms can be applied to the problem of identifying c-optimal experimental designs when there may be correlation between and within experimental units and evaluate the performance of relevant algorithms. We assume the data generating process is a generalised linear mixed model and show that the c-optimal design criterion is a monotone supermodular function amenable to a set of simple minimisation algorithms. We evaluate the performance of three relevant algorithms: the local search, the greedy search, and the reverse greedy search. We show that the local and reverse greedy searches provide comparable performance with the worst design outputs having variance $<10\%$ greater than the best design, across a range of covariance structures. We show that these algorithms perform as well or better than multiplicative methods that generate weights to place on experimental units. We extend these algorithms to identifying moole-robust c-optimal designs.
\end{abstract}

\noindent%
\vfill

\newpage
\spacingset{1.45} 
\section{Introduction}
\label{sec:intro}
We consider the question of how to identify a c-optimal design when the observations are correlated. In particular, we assume the data generating process can be described using a generalised linear mixed model (GLMM). For a $N \times 1$ vector of outcomes $y$, with an $N \times P$ matrix $X$ of covariates and a $N \times Q$ `design matrix' for random effects $Z$, a GLMM can be written as:
\begin{equation}
\label{eq:glmm}
    y \sim F(h^{-1}(X\beta + Z\mathbf{u}),\phi)
\end{equation}
where $\beta$ are mean function parameters, $h(.)$ is a link function, $F(.)$ is a distribution function with scale parameter(s) $\phi$, and $\mathbf{u} \sim N(0,D)$ is a vector of random effects with covariance matrix $D$. Such models provide a flexible parametric approach to estimation of covariate effects when the observations are correlated, such as longitudinal designs \citep{Zeger1988}, cluster randomised trials \citep{Hussey2007}, and geospatial statistical modelling \citep{Diggle1998}. 

The rows $i=1,...,N$ of matrices $X$ and $Z$ define a discrete set of possible observations in the experiment. In some cases, it is assumed that the discrete design points are formed by a uniform lattice over a continuous design space (e.g. \citet{Yang2013}). For many study designs with correlation, the observations are grouped into blocks, clusters, or other set; we refer to such a group as an `experimental unit'. An experimental unit is:
\begin{equation*}
    e_j \subset \{1,...,N\}
\end{equation*}
for $j=1,...,J$ where we assume $\bigcup_j e_j = \{1,...,N\}$ and $e_j \cap e_{j'} = \emptyset$ for $j\neq j'$, that is all observations are in one and only one experimental unit. Without loss of generality we assume that all the experimental units have the same size $r$. A \textit{design space} then consists of all the experimental units:
\begin{equation*}
    D := \{ e_j: j = 1,...,J\}
\end{equation*}
A \textit{design} $d \subset D$ of a maximum size $m<N$ is then:
\begin{equation*}
    d := \{ e_j: e_j \in D; |d| \leq m \}
\end{equation*}
The total number of observations in the design is $n=mr$. The design problem we face is then to find the design $d^* \subset D$ that minimises some objective function. There are multiple criteria used in the experimental design literature, here we only discuss c-optimality.

\subsection{Information Matrix}
We consider GLMMs where $F(.)$ is in the exponential family, including Gaussian, binomial, and Poisson models. The likelihood of the model parameters is 
\begin{equation}
\label{eq:lik1}
    L(\beta,\phi,\theta|y) = \int \prod_{i=1}^n f_{y|\mathbf{u}}(y_i|\mathbf{u}, \beta, \phi)f_{\mathbf{u}}(\mathbf{u}|\theta) d \mathbf{u}
\end{equation}
where $f_{y|\mathbf{u}}(y_i|\mathbf{u}, \beta, \phi) = \exp{(y_i\eta_i - c(\eta_i))/a(\phi) + d(y_i,\phi)}$, $f_{\mathbf{u}}$ is the multivariate Gaussian density, $\eta_i = x_i\beta + z_i\mathbf{u}$ is the linear predictor, $x_i$ and $z_i$ are the $i$th rows of matrices $X$ and $Z$, and $\theta$ are the parameters that define the covariance matrix $D$.

The information matrix for the model parameters $\beta$ is:
\begin{equation}
\label{eq:infomatpartial}
    M = X^T E_y\left[ \left( \frac{\partial \log L(\beta,\phi,\theta|y)  }{\partial \beta}\right) \left( \frac{\partial \log L(\beta,\phi,\theta|y)  }{\partial \beta}\right)^T |\beta, \theta, \phi \right] X
\end{equation}
For the generalised least squares estimator of the GLMM this information matrix is equivalent to
\begin{equation}
\label{eq:infomatgls}
    M = X^T \Sigma^{-1} X
\end{equation}
where $\Sigma = \text{Cov}(y)$. We use $M_d$ to denote the information matrix for design $d$.

The c-optimal objective function $g:2^{d} \to \mathbb{R}_{\geq 0}$ we consider is then:
\begin{equation}
\label{eq:funcg}
    g(d) = 
    \begin{cases}
      c^T M_d^{-1} c & \text{for } M_d \text{ positive semi-definite} \\
      \infty & \text{otherwise }
    \end{cases}
\end{equation}
where $c$ is a $p \times 1$ vector such that $c \in \text{range}(M)$ to ensure estimability of $c^T\beta$ \citep{Pukelsheim1980}. The vector $c$ often has all elements equal to zero except for a one at the position corresponding to the treatment effect parameter of interest. It is possible for a design not to produce a positive semi-definite information matrix, such as when the matrix $X$ is not of full rank, which will occur if there are too few observations, for example. For these cases we assume that the objective function takes the value of infinity, i.e. the design provides no information about the parameters. The consequences of this choice are discussed later. Our optimisation problem is the to find the design $d \subset D$ of size $m$ that minimises $g(d)$.

\subsection{Previous Literature}
It is generally not possible to identify exact c-optimal designs in this context. For designs with correlated observations perhaps the most common approximate method is to identify the optimal `weights' to place on each experimental unit or design point. These weights can be interpreted as an ``amount of effort'' to place on the observations in the design. Elfving's Theorem is a classic result in theory of optimal designs that underlies this method \citep{Elfving1952,Ford1992,Studden2005}. For independent identically distributed observations (i.e. when $\Sigma = \sigma^2 I)$ we can write the information matrix as the sum $M = \sum_i x_i^T x_i$ for all observations $i=1,...,N$. We can place a probability measure over the observations $\rho = \{(\rho_i,x_i): i=1,...N; \rho_k \in [0,1]\}$ where $\rho_i$ are the weights on each observation. The information matrix of the approximate design is $M(\rho) = \sum_i x_i^T x_i \rho_i$. Elfving's theorem provides a geometric characterisation of c-optimality and shows that an optimal design $\rho$ lies at the intersection of the convex hull of the $x_i$ and the vector defined by $c$. The values of $\rho$ for the optimal design can then be obtained using linear programming methods \citep{Harman2008}.

\citet{Holland-Letz2011} and  \citet{Sagnol2011} extended Elfving's theorem to design spaces with experimental units in which the observations may be correlated, but where there is no correlation \textit{between} experimental units. In this case, the information matrix (\ref{eq:infomatgls}) can be written as the sum of the information matrices for each experimental unit:
\begin{equation}
\label{eq:infomat}
    M = \sum_{j=1}^J X_{e_j}^T \Sigma_{e_j}^{-1} X_{e_j}
\end{equation}
where we use $X_{e_j}$ to represent the rows of $X$ in $e_j$, and similarly $\Sigma_{e_j}$ is the submatrix of $\Sigma$ given by the rows and columns in $e_j$. This expression can be rewritten as $\sum_j X_{e_j}^T F F^T X_{e_j}$ where $F_j$ is a square root of $\Sigma_{e_j}^{-1}$. The weights $\rho$ are placed on each experimental unit so that a design can be represented by the pairs $\{(e_1,\rho_1),...,(e_J,\rho_j)\}$ and the problem reduces to identifying the optimal weights using the generalised Elfving theorem. \citet{Sagnol2011} shows that the optimal weights for each experimental condition can be solved using conic opimisation methods with a second order cone program. We refer to these approaches generally as `multiplicative'. \citet{HollandLetz2012} provide an estimate on the lower bound of the efficiency of multiplicative methods in a article examining optimal assignment of individuals to different dose schedules in a pharmacokinetic study. Prior approaches to optimal experimental designs with correlated observations relied on asymptotic arguments \citep{Sacks1968,Muller2003,Nather1985}. 

One potential limitation of multiplicative methods is that there are several different approaches to rounding weights to integer totals of experimental units \citep{Balinski2002}. \citet{Pukelsheim1992} determine the optimal rounding scheme when at least one of each type of experimental unit is required. However, for many design problems this restriction is not necessary. As such, different rounding methods may produce different designs, which may not necessarily be optimal. Multiplicative methods also have the limitation that they cannot be extended to designs where there may be correlation \textit{between} experimental units. We may also wish to accommodate restrictions in the design space, such as a maximum or minimum number (or weight) on particular experimental conditions given practical restrictions. 

In this article, we show that `combinatorial' algorithms are applicable to the problem of identifying c-optimal experimental design with correlated observations both within and between experimental units and compare their performance on a set of example study designs. `Combinatorial' optimisation methods aim to select the optimal set of discrete items from a larger finite set. Where relevant, we refer to these methods as `combinatorial algorithms' to differentiate them from the multiplicative methods. These algorithms can identify local minima, however a combinatorial approach cannot guarantee a global minimum is found. Results from the optimisation literature show that the difference between the solutions from the algorithms and global minima can be bounded though. We discuss the relevant combinatorial algorithms in Section 2 and how they can be applied to the c-optimal design problem, including both Gaussian and non-Gaussian models. In Section 3 we compare the performance of these algorithms across a set of example problems. Section 4 compares the performance of multiplicative and combinatorial approaches, and section 5 extends the discussion to robust optimisation.

\section{Monotone Supermodular Function Minimisaton}
A function $g$ is called \textit{supermodular} if:
\begin{equation}
\label{eq:super}
    g(d \cup \{ e_j \}) - g(d) \geq g(d' \cup \{ e_j \}) - g(d')
\end{equation}
for all $d' \subseteq d$. That is, there are diminimising marginal reductions in the function with increasing size of the design. The function is \textit{monotone decreasing} if $d' \subseteq d \rightarrow g(d') \geq g(d)$

Equation (\ref{eq:infomat}) shows that when the observations in different experimental units are independent, then the information matrix can be written as a sum of information matrices for each unit. We can derive a more general expression for the marginal change to the information matrix when observations are added. Let $d'$ and $d$  be two designs such that $d' \subset d \subset D$ and $d = d' \cup d''$. We let $X_1$ and $X_2$ be the covariate matrices for designs $d'$ and $d''$, respectively, and $\Sigma_1$ and $\Sigma_2$ be their covariance matrices. $\Sigma_{12}$ is the covariance between the observations in designs $d'$ and $d''$. Then, 
\begin{align}
\label{eq:infomat2}
\begin{split}
    M_d &= \begin{bmatrix}
    X_1 \\
    X_2
    \end{bmatrix}^T
    \begin{bmatrix}
    \Sigma_1 & \Sigma_{12} \\
    \Sigma_{12}^T & \Sigma_{2}
    \end{bmatrix}^{-1}
    \begin{bmatrix}
    X_1 \\
    X_2
    \end{bmatrix} \\
    &= \begin{bmatrix}
    X_1 \\
    X_2
    \end{bmatrix}^T
    \begin{bmatrix}
    \Sigma_1^{-1} - \Sigma_1^{-1}\Sigma_{12}^{T}S^{-1}\Sigma_{12}\Sigma_1^{-1}  & -\Sigma_1^{-1}\Sigma_{12}^{T}S^{-1} \\
    -S^{-1}\Sigma_{12}\Sigma_1^{-1}& S^{-1}
    \end{bmatrix}
    \begin{bmatrix}
    X_1 \\
    X_2
    \end{bmatrix} \\
    &= X_1^T \Sigma^{-1}X_1 + X_2^TS^{-1}X_2 - X_1^T\Sigma_1^{-1}\Sigma_{12}^TS^{-1}X_2 - X_2^TS^{-1}\Sigma_{12}\Sigma_1^{-1}X_1 + \\
    & X_1^T\Sigma_1^{-1}\Sigma_{12}^TS^{-1}\Sigma_{12}\Sigma_1^{-1}X_1 \\
    &= M_{d'} + [X_2 - \Sigma_{12}^T\Sigma_1^{-1}X_1]^TS^{-1}[X_2 - \Sigma_{12}^T\Sigma_1^{-1}X_1] \\
    &= M_{d'} + \delta M_{d',d''}
    \end{split}
\end{align}
where $S = (\Sigma_2 - \Sigma_{12}^T\Sigma_1^{-1}\Sigma_{12})$ is the Schur complement, which we assume is invertible, and we use $\delta M_{d',d''}$ to represent the marginal change in the information matrix of $M_{d'}$ when the additional observations in $d''$ are added. It is evident that if $\Sigma_{12} = 0$ then $\delta M_{d',d''}$ reduces to $X_2^T\Sigma_2^{-1}X_2$ as in Equation (\ref{eq:infomat}). We also note that $\delta M_{d',d''}$ is positive definite such that $M_d \succeq M_{d'}$ (where $A \succeq B$ means $A-B$ is positive semi-definite), which implies that $g(d) \leq g(d')$ for $d' \subseteq d$. Thus, the function $g$ is monotone decreasing.

For our specific c-optimality problem, we can re-express (\ref{eq:super}) using Equation (\ref{eq:infomat2}) as:
\begin{align}
\begin{split}
\label{eq:schur}
    c^T([M_{d'} + \delta M_{d',d''} + \delta M_{d,e_j}]^{-1} & - [M_{d'} + \delta M_{d',d''}]^{-1})c  \geq 
     \\
    & c^T([M_{d'} + \delta M_{d',e_j}]^{-1}-M_{d'}^{-1})c  \\
    c^T[M_{d'} + \delta M_{d',d''} + (M_{d'} + \delta &M_{d',d''})\delta M_{d,e_j}^{-1}(M_{d'} + \delta M_{d',d''})]^{-1}c \leq \\
    & c^T[M_{d'} + M_{d'}\delta M_{d',e_j}^{-1}M_{d'}]^{-1}c
    \end{split}
\end{align}
where the second line follows from Hua's identity. The function is then supermodular if:
\begin{align}
\begin{split}
    M_{d}[ M_{d}^{-1} + \delta M_{d,e_j}^{-1} ]M_{d}  \succeq M_{d'}[ M_{d'}^{-1} + \delta M_{d',e_j}^{-1} ]M_{d'} 
    \end{split}
\end{align}
It can be shown that this condition is satisfied if $\delta M_{d,e_j}$ is symmetric, positive semidefinite for all $d$, which Equation (\ref{eq:schur}) shows to be the case if the covariance matrix and Schur complement are invertible. Thus, the c-optimal design problem for the GLMM under the GLS estimator is monotone supermodular.

A function is called \textit{submodular} if in Equation (\ref{eq:super}) the inequality is reversed. The maximisation of a submodular function is generally equivalent to the minimisation of a supermodular function \citep{Sviridenko2017}. The literature often discusses the former, but both problems have the same algorithms associated with finding approximate optimal solutions.


\subsection{Algorithms}
\subsubsection{Local search algorithm}
Algorithm 1 shows the \textit{local search} algorithm. We start with a random design of size $m$ and at each step of the algorithm the best swap of an experimental unit in the design with one not in the design is made until there are no more swaps that improve the design.

For monotone supermodular function minimisation in general, there is no guarantee the local search will converge to the globally optimal design. However, the algorithm does have a provable `constant factor approximation'. If $d$ is the output of the local search algorithm and $d^*$ is the global minimiser of the function, then the constant factor approximation is the upper bound of $g(d)/g(d^*)$. \citet{Fisher1978} showed that under a cardinality constraint (such as $|d| \leq n$) the optimality bound is $3/2$. \citet{Filmus2014} improved this bound to $(1+1/e)$ by using an auxilliary function in place of $g$ that excludes poor local optima. \citet{Feige1998} showed that further improving these bounds is an NP-hard problem and \citet{Nemhauser1978} showed that algorithms that improve on these bounds require an exponential number of function evaluations, rather than the polynomial number of the local search. In practice the local algorithm, and those discussed below, may perform significantly better than their lower bounds would suggest, however there exists little empirical evidence for the types of design we consider. We also note that \citet{Fedorov1972} developed the first local search algorithm for D-optimal designs (i.e. a design that maximises $\text{det}(M_d)$) with independent observations; several variants were later proposed \citep{Nguyen1992}. Although there is presently no proof that such an approach converges to a D-optimal design.

\begin{algorithm}
 \KwData{$X$,$Z$,$\Omega$,$\beta$,$\phi$,$D$}
 \KwResult{An optimal design $d^*$}
 Let $d_0$ be size $m$ design \;
 Set $\delta = 1$ and $d \leftarrow d_0$ \;
 \While{$\delta < 0$}{
 \lForEach{$e_j \in d$ and $e_{j'}\in D / d$}{
    Calculate $g(d / \{e_j\} \cup \{e_{j'}\})$; }
 Set $d' \leftarrow \argmin_{j,j'} g(d / \{e_j\} \cup \{e_{j'}\})$ \;
 $\delta = g(d') - g(d)$ \;
 \If{$\delta > 0$ }{
    $d \leftarrow d'$}
 }
 \caption{Local search algorithm}
\end{algorithm}

\subsubsection{Greedy and reverse greedy search algorithms}
Algorithm 2 shows the ``greedy algorithm''. We start from the empty set ($d=\emptyset$) and at each step of the algorithm add the experimental unit with the smallest marginal increase in the objective function. Rather than sequentially adding experimental units, one can start from the complete design space $D$ and sequentially remove units. This is the ``reverse greedy algorithm'', which is shown in Algorithm 3. 

\begin{algorithm}
 \KwData{$X$,$Z$,$\Omega$,$\beta$,$\phi$,$D$,n}
 \KwResult{An optimal design $d^*$}
 Let $d$ be a non-degenerate design of size $ p \leq s < m$ \;
 Set $k = 0$\;
 \While{$k < m$}{
 \lForEach{$e_{j}\in D / d$}{
    Calculate $g(d \cup \{e_{j}'\})$; }
 Set $d \leftarrow d \cup \argmin_{e_j} g(d \cup \{e_j\})$ \;
 $k \leftarrow k + 1$
 }
 \caption{Greedy search algorithm}
\end{algorithm}

\begin{algorithm}
 \KwData{$X$,$Z$,$\Omega$,$\beta$,$\phi$,$D$,n}
 \KwResult{An optimal design $d^*$}
 Let $d=D$ be the design containing all experimental conditions\;
 Set $k = J$\;
 \While{$k > m$}{
 \lForEach{$e_{j}\in d$}{
    Calculate $g(d / \{e_{j}'\})$; }
 Set $d \leftarrow d / \argmin_{e_j} g(d / \{e_j\})$ \;
 $k \leftarrow k - 1$
 }
 \caption{Reverse greedy search algorithm}
\end{algorithm}

The constant factor approximations for the greedy and reverse greedy algorithms are more complex that the local search case. In the case of submodular function maximisation, a famous result is that the constant factor approximation is $1+1/e$ \citep{Nemhauser1978}. However, this result does not carry over to minimising a supermodular function \citep{Ilev2001}. Indeed, it is not possible to implement the greedy algorithm for the design problems we discuss, as all designs with fewer than $p$ observations, and many with more than $p$ observations, will result in a non-positive semidefinite information matrix. We can start the algorithm from a random small design, as Algorithm 2 describes, but this of course would sacrifice any theoretical guarantees. \citet{Ilev2001} discusses the approximation factor for the reverse greedy algorithm in the case of minimising a supermodular function. The result depends on the `steepness' or curvature of the function $g$, which is defined as:
\begin{equation*}
    \max_{e \in D}\frac{(g(e)-g(\emptyset)) - (g(D) - g(D/e))}{(g(e)-g(\emptyset)) }
\end{equation*}
In Equation (\ref{eq:funcg}) we specified that the function had infinite variance for the empty set, in which case the steepness would be one, which is equivalent to an unbounded curvature. In these cases the reverse greedy search does not have an approximation factor \citep{Ilev2001,Sviridenko2017}. Specifying the value of the function to be undefined would also fail to provide a bound.

Greedy algorithms have been used in the experimental design literature previously. \citet{Yang2013} developed a continuous sequential/greedy algorithm to identify optimal designs for generalised linear models under a range optimality criteria, although not for c-optimality. They discretised the design space using a regular lattice and showed convergence to optimal designs as the number of lattice cells grows. Variants and combinations of these methods have also been proposed, for example, the `Cocktail algorithm' combines a sequential algorithm with two other algorithms in each step to find D-optimal designs \citep{Yu2011}. \citet{Fedorov1972} proposed a variant of this algorithm for D-optimal designs, often called a sequential algorithm, in which observations are sequentially added to an existing design until a convergence criterion identifying D-optimality is reached. \citet{Fedorov1972} showed this algorithm produced a D-optimal design for linear models without cardinality constraint. Accelerated (or adaptive) greedy algorithms provide significant computational improvements on the standard greedy algorithm by avoiding recomputation of the objective function \citep{Robertazzi1989}. Accelerated greedy algorithms have been used in experimental design, designing sensor networks, and other problems \citep{Yang2019,Zou2016,Guo2019}. However, again, there are few applications for c-optimality.

Given the lack of theoretical guarantees, one may consider these algorithms irrelevant to the c-optimal design problem. However, for some of the areas we use as examples below, they have been used informally. For example, there has been growing interest in methods to identify c-optimal designs for cluster randomised trials (e.g. \citet{Girling2016,Hooper2020}). Several recent articles have used an algorithmic approach that involves sequential removal of observations from a design space to identify c-optimal cluster randomised trial designs \citep{Hooper2020}, or using the change in variance of treatment effect estimators when experimental units are removed to identify efficient designs \citep{Kasza2019}. While these algorithms lack theoretical guarantees, they may empirically still perform adequately for these design problems. They also run faster than the local search. So we include them in the empirical comparisons below.

\subsection{Computation and approximation}
\subsubsection{Information Matrix Approximations}
\label{sec:approximations}
The greatest limitation on executing these algorithms is the evaluation of the information matrix (\ref{eq:infomatgls}) as it requires estimation and inversion of $\Sigma$ or estimation of the gradient of the log likelihood. As we discuss in Section \ref{sec:algeff}, once the covariance matrix is obtained, updating its inverse after adding or removing an observation can be done relatively efficiently. However, an efficient means of generating $\Sigma$ is still required for non-linear models. \citet{Breslow1993} used the marginal quasilikelihood of the GLMM to propose the following first-order approximation:
\begin{equation}
\label{eq:sigapprox}
    \Sigma \approx W^{-1} + ZDZ^T
\end{equation}
where $W$ is a diagonal matrix with entries $W_{ii} = \left( \frac{\partial h^{-1}(\eta)}{\partial \eta}^2 \text{Var}(y|\mathbf{u}) \right)$, which are recognisable as the GLM iterated weights \citep{mccullagh2019generalized}. The approximation is exact for the Gaussian model with identity link. Higher order approximations exist in the literature, however, their use has not been found to improve the quality of optimal designs, at least in the case of D-optimality. \citet{Waite2015} consider the case of D-optimal designs and compare (\ref{eq:sigapprox}) with the GEE working covariance matrix. They find it does not perform as well as the approximation based on the marginal quasilikelihood.  We do not consider the GEE covariance in this article, as we aim to use explicit covariance functions with different parameterisations. 

\citet{Zeger1988} suggests that when using the marginal quasilikelihood a better approximation to the marginal mean can be found by ``attenuating'' the linear predictor in non-linear models. For example, with the log link the ``attenuated'' mean is $E(y_i) \approx h^{-1}(x_i\beta + z_iDz_i^T/2)$ and for the logit link  $E(y_i) \approx h^{-1}(x_i\beta\vert c Dz_i^Tz_i + I \vert^{-1/2})$ with $c = 16\sqrt{3}/(15\pi)$. \citet{Waite2015} find that using attenuated parameters with the approximation (\ref{eq:sigapprox}) can achieve more efficient designs, at least for D-optimality. For the non-linear models below we compare with and without attenutation. R\citet{Waite2015} also propose approximations based on (\ref{eq:infomatpartial}). They consider blocked designs where there is no correlation between experimental units, and so the information matrix can be computed as the sum of information matrices of the units in the design as Equation (\ref{eq:infomat2}). For a binomial-logistic model, the information matrix can then be calculated using (\ref{eq:infomatpartial}) by completely enumerating the outcome space if the size of the experimental unit is relatively small. However, we consider designs where there may be correlation between experimental units, limiting the computational tractability of such an approach as a complete enumeration of the outcome space would be infeasible.

\subsubsection{Algorithm Efficiency}
\label{sec:algeff}
Equation (\ref{eq:infomat2}) also shows how we can achieve some computational efficiency with correlated observations. A naive optimisation approach that recalculated the information matrix for each design would require at least $O(n^{3})$ operations by needing to invert the covariance matrix each time. We can iteratively add or remove single observations at a time, i.e. moving from $d$ to $d / \{i\} \cup \{i'\}$ and $i \neq i'$, so the calculation only requires $O(rn^{2})$ operations to add or remove an experimental condition through rank-1 up/down dates of the inverse covariance matrix (see supplementary material). The total running time of the local search algorithm scales with $O(m^4 r^3 (J-m))$, as we have to evaluate swapping $m$ experimental units with $J-m$ remaining units of size $r$ up to $m$ times. The experimental units are not always unique, in these cases we can detect any duplicated experimental conditions and only evaluate any swap involving it once, while keeping track of the number of copies in the design space and design, to reduce running time. As such the values of $m$ and $J$ in the expression for complexity can be interpreted as the numbers of \textit{unique} experimental units in the design and design space, respectively.

The computational complexity of the greedy search algorithm scales as $O(m^3r^3(J-m))$, however it generally runs much faster than the local search since most function evaluations are of designs smaller than $m$. The complexity of the reverse greedy algorithm scales as $O(J^3r^3(J-m))$.

\section{Comparative performance}
\subsection{Comparison of Algorithms}
We consider several examples to compare the three algorithms described above. We compare them in two areas: quality of solution and computational time. As the local and greedy search algorithms have a random starting set, we run them each 100 times; the reverse greedy search is deterministic and so is run only once per example. 

We use the approximation (\ref{eq:sigapprox}) for all the analyses. For the examples using a Gaussian-identity model, the approximation is exact. For non-linear models (binomial-logistic, binomial-log, and poisson-log) we make the additional comparison between an approximation with and without attenuation as described in Section \ref{sec:approximations}.

For each example we calculate the `relative efficiency' as the ratio of the variance (i.e. the value of $c^TM_d^{-1}c$) of the design(s) from each algorithm compared to the variance of the best design from all algorithms expressed as a percentage. For the non-linear models, we only evaluate the single best design from each algorithm, with and without attenuated parameters, using Equation (\ref{eq:infomat}). Enumerating the complete outcome space to evaluate the expectations in (\ref{eq:infomat}) would not be possible, so we use Monte Carlo integration with 100,000 iterations to estimate the relative variance.

We also report the approximate running time of each algorithm. Timings were made on a computer with Intel Core i7-9700K, 32GB RAM, Windows 10, and the program was compiled with gcc compiler. We make all these algorithms available as part of the \texttt{glmmrOptim} package for R.

\subsection{Applied Examples}
Our examples are derived from two study types that motivated this article. Identifying optimal cluster randomised trial designs, and determining optimal sampling locations to estimate treatment effects in a geospatial setting. We describe each of these in turn along with their associated examples.

\subsubsection{Cluster randomised trial}

\begin{figure}
    \centering
    \includegraphics[width=0.3\textwidth]{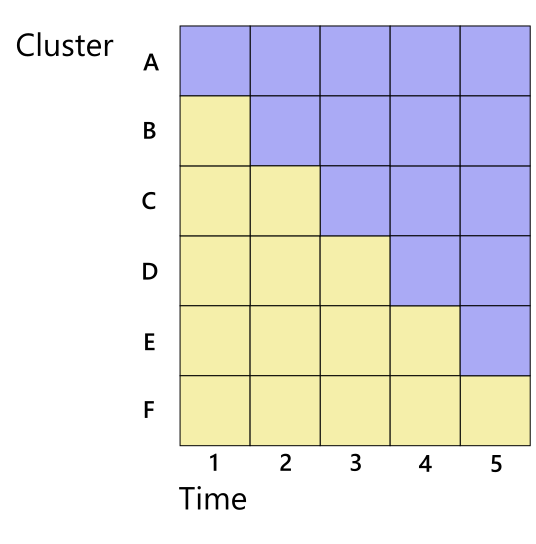}
    \caption{An illustration of a cluster randomised trial design space. Each cell represents a cluster-period and is comprised of observations from individuals within the cell. Yellow = control status with no intervention; blue = with intervention.}
    \label{fig:cdiag}
\end{figure}

A cluster randomised trial is a type of randomised trial design in which groups, or `clusters', of individuals are randomly allocated to receive either a treatment or control. Cluster trials are typically used to evaluate interventions that are applied to groups of people rather than individuals, for example, quality improvement initiatives for healthcare clinics, or educational interventions in classrooms. Figure \ref{fig:cdiag} describes a design space for a cluster randomised trial with repeated measures. Each cell is a cluster-period within which we can observe multiple individuals. There are $K=6$ clusters and $T=5$ total time periods. The linear predictor for an observation $i$ in cluster $k$ at time $t$ is:
\begin{equation} 
\label{eq:mfun1}
    \eta_{ikt} = \beta_0 \Delta_{kt} + \beta_1 \tau_t + \epsilon_{ikt}
\end{equation}
where $\Delta_{kt}$ is a treatment indicator equal to one if the cell has the intervention and zero otherwise as shown in Figure 1, $\tau_t$ are $T$ time-period indicators (so we do not include an intercept). We consider two covariance function specifications for $\epsilon_{ikt}$ at the cluster-level and examine both cross-sectional and cohort designs. The covariance functions represent the most widely used specifications for these studies \citep{Li2021}. First, an exchangeable covariance function with cluster and cluster-period random effects with cross-sectional sampling in each time period \citep{Hussey2007,Hemming2015b}: 
\begin{equation*}
\label{eq:cfun1}
    Cov(\epsilon_{ikt},\epsilon_{i'k't'}) = 
    \begin{cases}
      \sigma^2_1 + \sigma^2_2 & \text{ if } k=k',t=t'\\
      \sigma^2_1 & \text{ if } k=k',t\neq t'\\
      0 & \text{ otherwise}\\
    \end{cases}
\end{equation*}
Second, an autoregressive covariance function:
\begin{equation*}
\label{eq:cfunc2}
    Cov(\epsilon_{ikt},\epsilon_{i'k't'}) = 
    \begin{cases}
      \sigma^2_1\lambda^{|t-t'|} & \text{ if } k=k'\\
      0 & \text{ otherwise}\\
    \end{cases}
\end{equation*}

For models with cohort effects where the same individuals appear in each cluster in every period, we modify these covariance functions by adding an additional term $\sigma^2_c$ to the covariance if the individual is the same $i=i'$. For Gaussian models we notate the observation-level variance of the error term as $\sigma^2_e$.

Table \ref{tab:specs1} lists the different model specifications we examined for the cluster randomised trial and other examples. We include both linear and non-linear models with differing covariance structures and parameters. The choice of covariates generally represents a range from `high' to `low' levels of between-cluster correlation in these settings \citep{Hemming2015b,Hemming2020}. We specify a maximum number of individuals per cluster-period of 10, and aim to identify an optimal design of $m=100$ individuals (out of a possible 300); an experimental unit is a single observation from an individual. We set $c=(1,0,...,0)^T$.

\begin{table}\footnotesize
    \centering
    \begin{tabular}{lccccc}
         \toprule
         & \multicolumn{3}{c}{\textbf{Model}}  & \multicolumn{2}{c}{\textbf{Parameters}} \\
         & $F$ & $h$ & Cov &  $\beta$ & $\theta$ \\
         \midrule
         \multicolumn{6}{c}{\textit{Cluster trial - cross-sectional}}\\
         \midrule
         A & Gaussian & Identity & Exc. & N/A & $\sigma_1 = 0.25$ $\sigma_2 = 0.1$ $\sigma_e=1$\\ 
         B & Gaussian & Identity & Exc. & N/A & $\sigma_1 = 0.1$ $\sigma_2 = 0.1$$\sigma_e=1$ \\ 
         C & Gaussian & Identity & AR1 & N/A & $\sigma_1 = 0.25$ $\lambda = 0.6$$\sigma_e=1$ \\ 
         D & Gaussian & Identity & AR1 & N/A & $\sigma_1 = 0.1$ $\lambda=0.9$ $\sigma_e=1$\\ 
         E & Binomial & Logit & Exc. & $\beta_0 = 0.1$  & $\sigma_1 = 0.25$ $\sigma_2 = 0.1$ \\ 
         &&&& $\beta_1=[-0.5,-0.3,-0.1,0.1,0.3]$ & \\
         F & Binomial & Logit & Exc. & $\beta_0 = 0.1$  & $\sigma_1 = 0.1$ $\sigma_2 = 0.1$ \\ 
         &&&& $\beta_1=[-0.5,-0.3,-0.1,0.1,0.3]$ & \\
         G & Binomial & Log & AR1 & $\beta_0 = 0.1$  & $\sigma_1 = 0.25$ $\lambda = 0.6$ \\ 
         & & && $\beta_1=[-1.5,-1.3,-1.1,-0.9,-0.7]$ & \\
         H & Binomial & Log & AR1 & $\beta_0 = 0.1$  & $\sigma_1 = 0.1$ $\lambda=0.9$ \\
         & & && $\beta_1=[-1.5,-1.3,-1.1,-0.9,-0.7]$ & \\
         \midrule
         \multicolumn{6}{c}{\textit{Cluster trial - cohort}}\\
         \midrule
         I & Gaussian & Identity & Exc. & N/A & $\sigma_1 = 0.25$ $\sigma_2 = 0.1$ $\sigma^2_c = 0.8$ $\sigma^2_e = 0.2$\\
         J & Gaussian & Identity & Exc. & N/A & $\sigma_1 = 0.1$ $\sigma_2 = 0.1$ $\sigma^2_c = 0.8$ $\sigma^2_e = 0.2$\\ 
         K & Gaussian & Identity & AR1 & N/A & $\sigma_1 = 0.25$ $\lambda = 0.6$ $\sigma^2_c = 0.8$ $\sigma^2_e = 0.2$\\ 
         L & Gaussian & Identity & AR1 & N/A & $\sigma_1 = 0.1$ $\lambda=0.9$ $\sigma^2_c = 0.8$ $\sigma^2_e = 0.2$\\ 
         \midrule
         \multicolumn{6}{c}{\textit{Geospatial sampling}}\\
         \midrule
         M & Gaussian & Identity & Exp. & N/A & $\sigma_1 = 0.25$ $\lambda=0.25$ \\ 
         N & Binomial & Logit & Exp. & $\beta_0 = 0$ $\beta_1=\log(2)$, $\beta_2 = 4$ & $\sigma_1 = 0.25$ $\lambda=0.25$ \\
         \bottomrule
    \end{tabular}
    \caption{Model specifications for the comparison examples for the cluster randomised trial. Exc. = exchangeable covariance function, AR1 = autoregressive covariance function, Exp. = exponential covariance function.}
    \label{tab:specs1}
\end{table}

\subsubsection{Geospatial sampling}
There is a broad literature on selecting the optimal sampling locations (and times) to draw samples across an area of interest (e.g. \citet{Chipeta2017}). However, these sampling patterns are generally designed to estimate a statistic like the prevalence of a disease across an area and its spatio-temporal distribution. A related, but possibly more complex, question asks where across an area one should sample to provide the most efficient estimates of point-source interventions with spatially-heterogeneous effects. A geospatial statistical model can be represented as a GLMM \citep{Diggle1998}, thus if the possible sampling locations are discretised, the design problem is amenable to the methods in this article.

Our design space is a unit-square $A = [0,1]^2$. The space is divided into a regular $15 \times 15$ lattice, where observations are made at the cell centroids $a \in A$. An intervention is located at the point $z = (0.5,0.5)$. The mean function is specified as:
\begin{equation}
\label{eq:mfun3}
    \eta_i(a) = \beta_0 + \beta_1\text{exp}(-\beta_2 |a-z|) + \epsilon_{a}
\end{equation} 
To accomodate the non-linear mean function in the framework described above we use an additional first-order approximation to the information matrix (following \citet{Holland-Letz2011,HollandLetz2012} and others):
\begin{equation}
    M_d = F^T \Sigma^{-1} F
\end{equation}
where the first column of $F$ is a vector of ones, the second column is $\partial \mu/ \partial \beta_1 = \text{exp}(-\beta_2 |a-z|)$, and the third column is $\partial \mu/ \partial \beta_2 = -\beta_1 |a-z|\text{exp}(-\beta_2 |a-z|)$. We set $\beta_0 = 1$, $\beta_1 = \text{ln}(2)$, and $\beta_2 = 4$. We specify a Poisson distribution with log link function. Finally, we specify an exponential covariance function:
\begin{equation}
    Cov(\epsilon_{a},\epsilon_{a'}) = \sigma^2_1\text{exp}(-\lambda|a-a'|)
\end{equation}
which is commonly used in geospatial applications. Our aim is to find an optimal design of size $m=80$ (of a total possible 325). We set $c=(0,1,0.1)^T$. Table \ref{tab:specs1} lists the parameter and model specifications for these examples.

\subsection{Results}

\begin{table}
    \centering
    \begin{tabular}{l|cccccc}
    \toprule   
     & \multicolumn{2}{c}{\textbf{Local}} & \multicolumn{2}{c}{\textbf{Greedy}} & \multicolumn{2}{c}{\textbf{Reverse greedy}} \\
     & \textit{Rel.eff.} & \textit{Time (s)} & \textit{Rel.eff.} & \textit{Time (s)} & \textit{Rel.eff.} & \textit{Time (s)} \\
    \midrule \midrule
       A  & 100.0 - 100.2 & 1 - 2 & 102.9 - 109.1 & 0 - 1 & 100.0 & 2  \\
       B  & 100.0 - 100.4 & 1 - 2 & 102.2 - 109.5 & 0 - 1 & 100.0 & 2  \\
       C  & 100.0 - 100.2 & 1 - 2 & 101.4 - 107.0 & 0 - 1 & 100.1 & 2  \\
       D  & 100.0 - 100.8 & 1 - 2 & 103.9 - 109.5 & 0 - 1 & 100.0 & 2  \\
       \midrule \midrule
       I & 101.2 - 108.2 & 40 - 50 & 116.0 - 345.6 & 1 - 2 & 100.0 & 10 \\
       J & 100.3 - 106.0 & 40 - 50  & 114.0 - 201.0 & 1 - 2 & 100.0 & 10\\
       K & 100.9 - 106.2 & 40 - 50 & 111.8 - 184.1 & 1 -2 & 100.0 & 10 \\
       L & 101.9 - 112.7 & 40 - 50 & 130.3 - 161.3 & 1- 2 & 100.0 & 10 \\
       \midrule \midrule
       M & 100.0 - 100.0 & 90 - 100 & 100.3 - 101.9 & 2 - 3 & 100.0 & 3 \\
       \bottomrule
    \end{tabular}
    \caption{Relative efficiency and approximate running time of the three algorithms for Examples using Gaussian-Identity model.}
    \label{tab:res1}
\end{table}

\begin{figure}
    \centering
    \includegraphics[width=\textwidth]{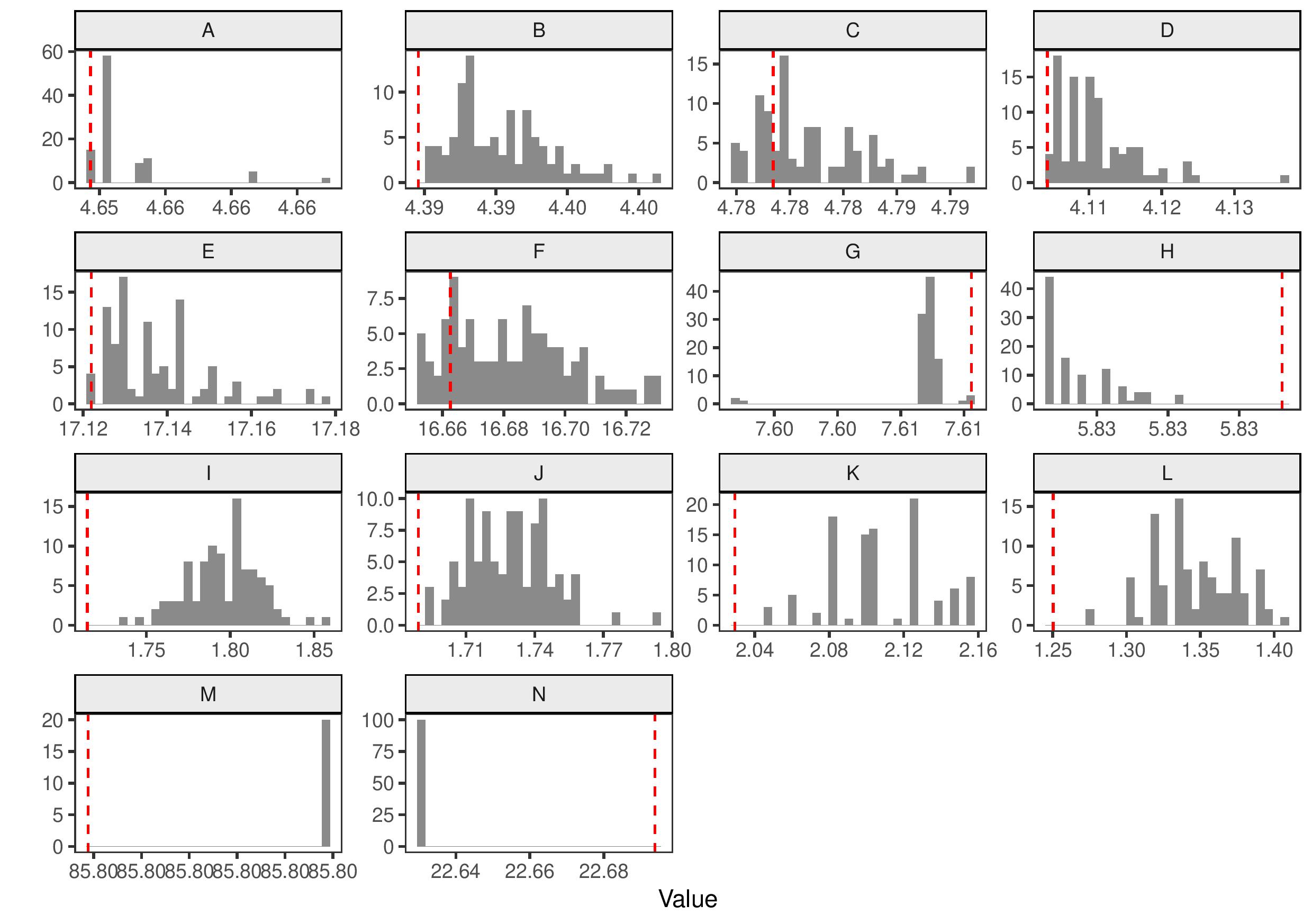}
    \caption{Histogram showing the values of the c-optimal criterion from the local search for each of the examples. The dashed red line indicates the value from the reverse greedy search. x-axis values are multiplied by 100.}
    \label{fig:histlocal}
\end{figure}

Table \ref{tab:res1} reports the relative efficiency and approximate computational times for the Gaussian examples. The greedy search performed the worst of the three algorithms is all examples. The local and reverse greedy searches both often found the optimal design, although the reverse greedy algorithm was more consistent with the worst design having a variance only 0.1\% greater than the local search. Figure \ref{fig:histlocal} shows the distribution of variances of the designs from the local search compared to the reverse greedy search. For the cross-sectional cluster trial designs the worst design produced by the local search had a variance less that 1\% greater than the best design, but with the cohort cluster trial designs this rose to 12\%. The same design was produced on every iteration for the geospatial example. Running times for these examples ranged from 1 to 100 seconds. The local search scaled poorly. The cross-sectional cluster trial examples had 30 unique experimental units compared to 60 for the cohort design and 325 for the geospatial example.

\begin{figure}
    \centering
    \includegraphics[width=\textwidth]{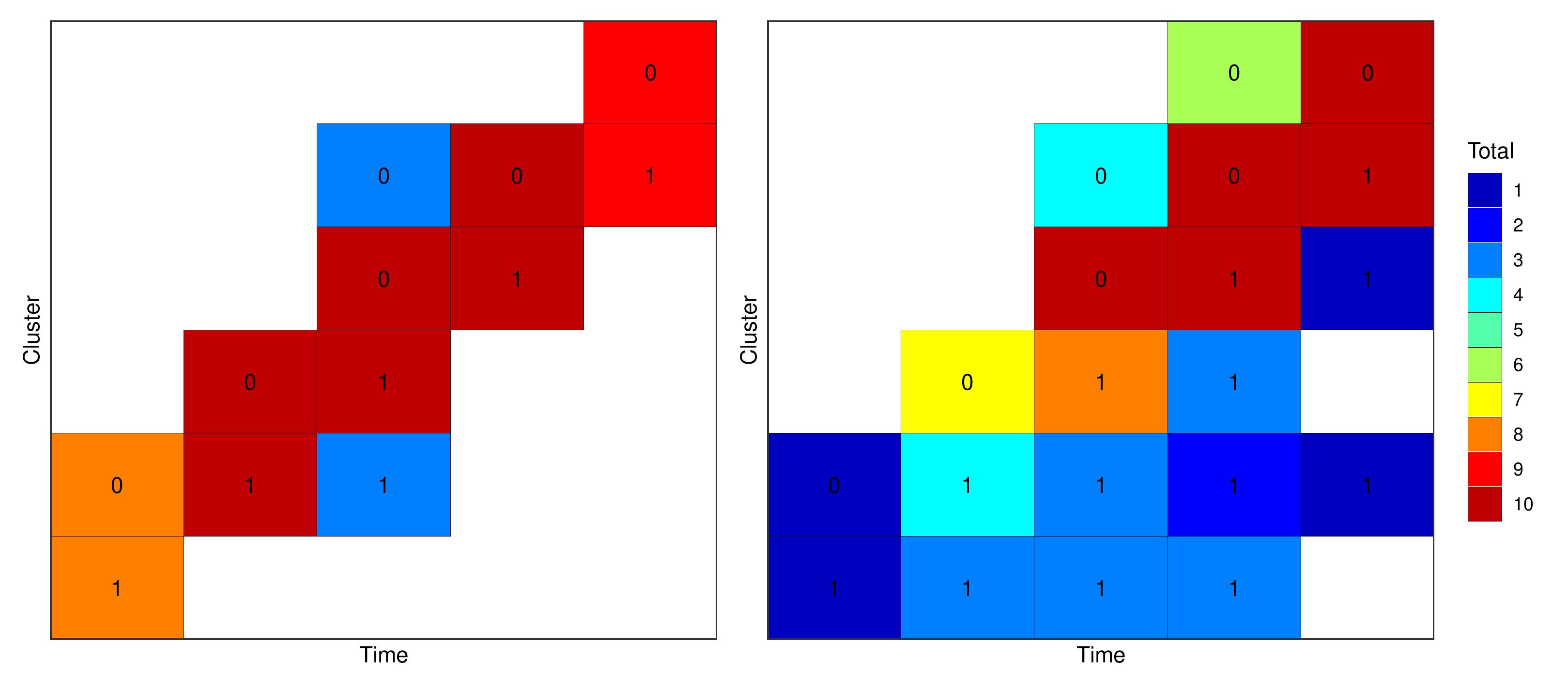}
    \caption{Optimal designs for Examples E and G. The 0 and 1 labels in the top row indicate the treatment status of the cluster period.}
    \label{fig:optdes}
\end{figure}

\begin{table}
    \centering
    \begin{tabular}{lc|cccccc}
    \toprule
     & Att. & \multicolumn{2}{c}{\textbf{Local}} & \multicolumn{2}{c}{\textbf{Greedy}} & \multicolumn{2}{c}{\textbf{Reverse greedy}} \\
     && \textit{Rel.eff.} & \textit{Time (s)} & \textit{Rel.eff.} & \textit{Time (s)} & \textit{Rel.eff.} & \textit{Time (s)} \\
    \midrule \midrule
       \multirow{2}{*}{E}  & N & 100.0 & 1 - 2 & 104.4 & 0 - 1 & 100.0 & 2  \\
         & Y & 100.7 & 1 - 2 & 101.1 & 0 - 1 & 100.0 & 2 \\
       \midrule
       \multirow{2}{*}{F}  & N & 100.0 & 1 - 2 & 101.9 & 0 - 1 & 100.0 & 2 \\
         & Y & 100.0 & 1 - 2 & 100.0 & 0 - 1 & 100.0 & 2 \\
       \midrule
       \multirow{2}{*}{G}  & N & 100.0 & 1 - 2 & 100.4 & 0 - 1 & 100.2 & 2 \\
        & Y & 100.0 & 1 - 2 & 100.3 & 0 - 1 & 100.2 & 2 \\
       \midrule
       \multirow{2}{*}{H} & N & 100.0 & 1 - 2 & 100.5 & 0 - 1 & 100.0 & 2 \\
        & Y & 100.0 & 1 - 2 & 100.5 & 0 - 1 & 100.0 & 2\\
        \midrule \midrule
       \multirow{2}{*}{N} & N & 100.0 & 90 - 100 & 100.2 & 2 - 3 & 100.0 & 3 \\
        & Y & 100.0 & 90 - 100 & 100.1 & 2 - 3 & 100.0 & 3 \\
       \bottomrule
    \end{tabular}
    \caption{Relative efficiency and approximate running of the best designs from the three algorithms with and without attenutation (Att.) for the examples with non-Gaussian models.}
    \label{tab:res2}
\end{table}

\begin{figure}
    \centering
    \includegraphics[width=0.75\textwidth]{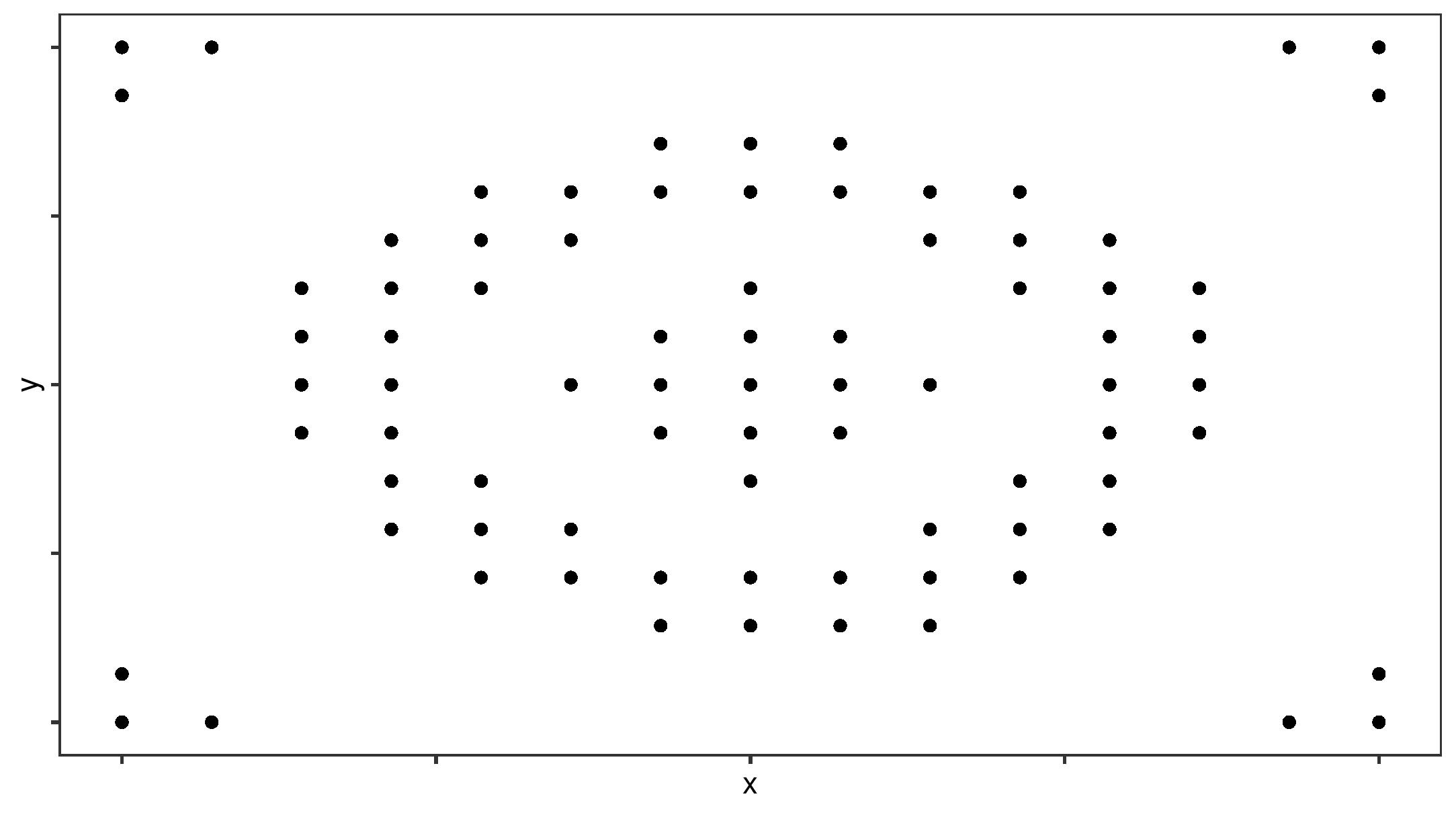}
    \caption{Optimal geospatial sampling pattern for Example I.}
    \label{fig:optdes2}
\end{figure}

Table \ref{tab:res2} reports the results for the non-Gaussian examples. Similarly to the Gaussian examples, the local and reverse greedy searches performed the best with either one or both finding the best design. Attenuation appears to make little difference to the quality of the solutions in these examples. Figures \ref{fig:optdes} and \ref{fig:optdes2} show the best design for examples E, H, and M. Notably, the optimal design for Example H is not symmetric owing to the heteroskedastic variance of the binomial-log model.

\section{Comparison with Multiplicative Methods}
An approximate `multiplicative' design is characterised by a probability measure over unique experimental units $\phi = \{(\rho_j,e_j): j=1,...J; \rho_j \in [0,1]; \sum_j \rho_j =1 \}$ where $\rho_j$ are the weights associated with each experimental units. These weights can be identified for design spaces with uncorrelated experimental units \citep{HollandLetz2012,Sagnol2011}. However, there are several methods for rounding proportions to integer counts that sum to a given total \citep{Balinski2002}. Briefly, for a given set of weights $\rho_j$ and a target total of number of experimental units $m$ in the design, the methods for determining the number of each experimental units $m_j$ such that $\sum_j m_j = m$, are:
\begin{enumerate}
    \item \textit{Hamilton's method} Let $\pi_j = n*\rho_j$, we assign $\lfloor \pi_j \rfloor$ of each experimental condition. The remaining total is filled by the experimental conditions with the largest remainders $|\pi_j - \lfloor \pi_j \rfloor |$ until we have $n$ experimental conditions.
    \item \textit{Divisor methods} Start with all $n_j = 0$ and let $\pi_j = n*\phi_j$. Proceeding iteratively, we choose the next experimental condition in the design to be that with $\max_j \pi_j/\alpha(n_j)$, for which we update the total until the condition $\sum_j n_j = n$ is met.
    \begin{enumerate}
        \item \textit{Jefferson's method} $\alpha(n_j) = n_j + 1$
        \item \textit{Webster's method} $\alpha(n_j) = n_j + 0.5$
        \item \textit{Adam's method} $\alpha(n_j) = n_j$. Initially we include one of each experimental condition with $\pi_j > 0$.
    \end{enumerate}
\end{enumerate}
\citet{Pukelsheim1992} showed that the optimal rounding method was a variant of Adam's method, although under the assumption that there is at least one of each experimental unit. However, for many experimental design problems this assumption is not required, including the example we examine below. To compare the performance of the multiplicative methods discussed in Section 1.2, we take the best design from the three combinatorial algorithms, and compare it to the design with the lowest variance from across all rounding methods. To identify the optimal approximate design $\rho$ we use the second-order cone program proposed by \citet{Sagnol2011}, which is also implemented in the R package \texttt{glmmrOptim} along with the rounding methods.

\subsection{Example}
We return to the cluster randomised trial examples, setting the experimental unit to a whole cluster sequence, i.e. a row consisting of five time periods each with ten individuals in Figure \ref{fig:cdiag}. We wish to include $m$ clusters in the design, each of which may be assigned to any of the experimental conditions, and consider $m=6$ to $m=30$. We use the models specified as Examples A-D in Table \ref{tab:specs1}. 

\begin{figure}
    \centering
    \includegraphics[width=\textwidth]{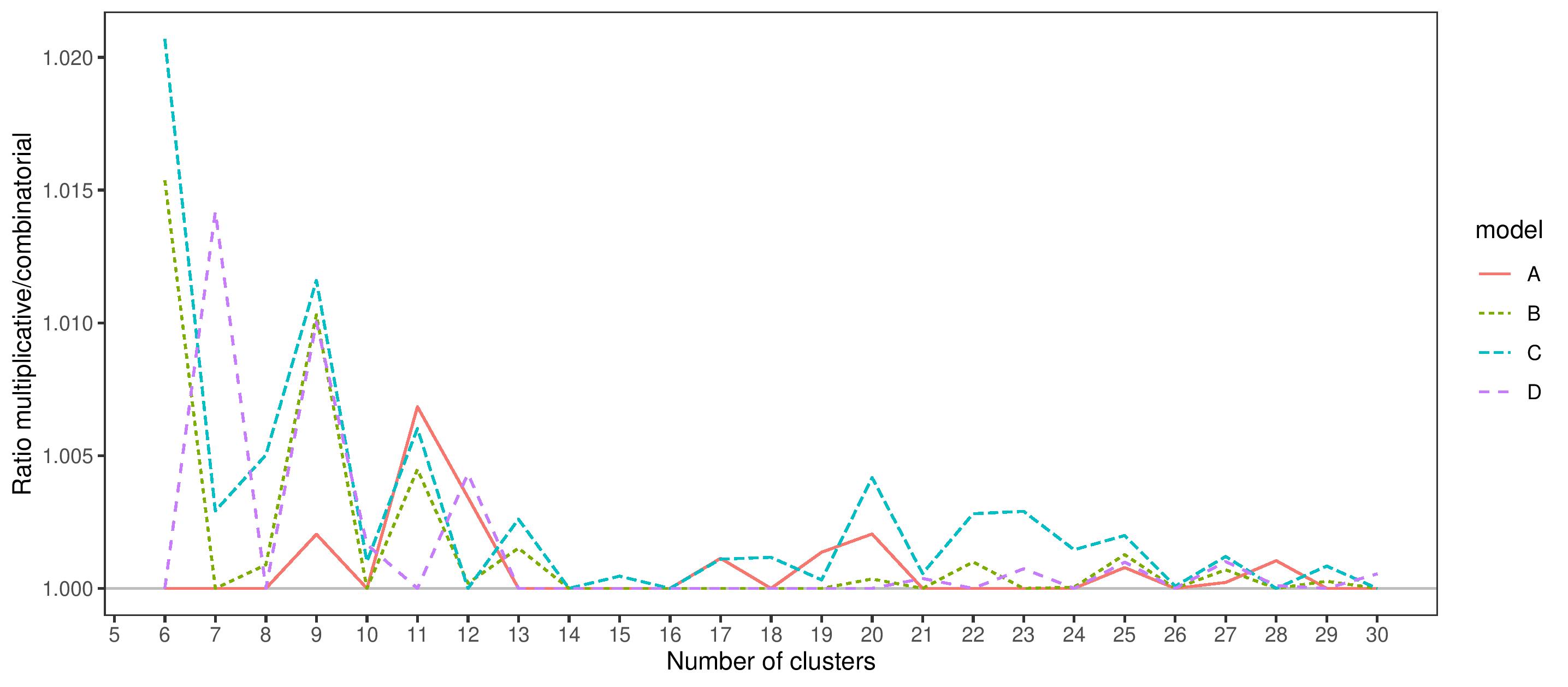}
    \caption{Ratio of the variance of the design from each method to the highest variance for 6 to 30 clusters in example \textit{D}}
    \label{fig:approx}
\end{figure}

Figure \ref{fig:approx} presents the ratio of the variance of the best design from the multiplicative algorithm to the variance of the best design from the combinatorial algorithms. For smaller sample sizes, particularly with an odd sample size, the multiplicative algorithm performed worst than the comibinatorial approach, although the worst design had a variance only 2\% larger than the best design. For larger sample sizes the two methods generally produced the same optimal design.

\section{Robust optimal designs}
The analysis and discussion so far has been based on the assumption that the correct model specification is known. However, this is typically an unrealistic assumption, and it is well known that an optimal design for one model may perform very poorly for an alternative model. We consider a model robust optimal design criterion that is amenable to combinatorial optimisation. Following \citet{Dette1993}, we assume that the ``true'' model belongs to a known class of GLMMs. We can define a model with the collection $G = (F,h,\beta,\theta)$. The class of models is then
\begin{equation}
    \mathcal{G}_U = \{G_1,...,G_U \}
\end{equation}
The vector $\rho = (\rho_1,...,\rho_U)$ with $\rho > 0$ is then called a prior for the class $\mathcal{G}$ with the values reflecting the belief about the relative probability or adequacy of each model. The objective function (\ref{eq:funcg}) can be written for a specific model as $g_u(d) = g(d;G_u)$ and we can define the robust objective function as:
\begin{equation}
\label{eq:robfunc}
    h(d) = \sum_{u=1}^U \rho_u g_u(d)
\end{equation}
The objective function (\ref{eq:robfunc}) is also monotone supermodular since $\sum_{u=1}^U \rho_u g_h(d \cup \{e \}) - \sum_{u=1}^U \rho_u g_h(d) = \sum_{u=1}^U \rho_u [g_h(d \cup \{e \}) - g_h(d)] \leq \sum_{u=1}^U \rho_u [g_h(d' \cup \{e \}) - g_h(d')]$ and $\sum_{u=1}^U \rho_u g_h(d) \geq \sum_{u=1}^U \rho_u g_h(d')$ for $d' \subseteq d$ if all the $g_u$ are themselves monotone submodular. We can therefore use the algorithms described above. A design minimising (\ref{eq:robfunc}) is then said to be optimal for $\mathcal{G}_U$ over the prior $\rho$. We note that other robust specifications such as minimax ($h(d) = \max_u g_u(d)$) are not supermodular, so we do not consider them here.

\citet{Dette1993} provides a geometric characterisation of the model robust criterion for c-optimal designs with uncorrelated observations, using the objective function $h(d) = \sum_u \rho_u \log(g_u(d))$, building on similar work for D-optimal designs. However, this has not yet been extended to correlated experimental units to permit use of multiplicative methods in this context.

We examine two examples for the robust optimal design. The examples A-D and E-H are taken as two classes of models. We assign equal weight to each design in each class as the prior. For the class E-H, we use both attenuated and non-attenutated linear predictors for the approximation.

\subsection{Results}
\begin{figure}
    \centering
    \includegraphics[width=\textwidth]{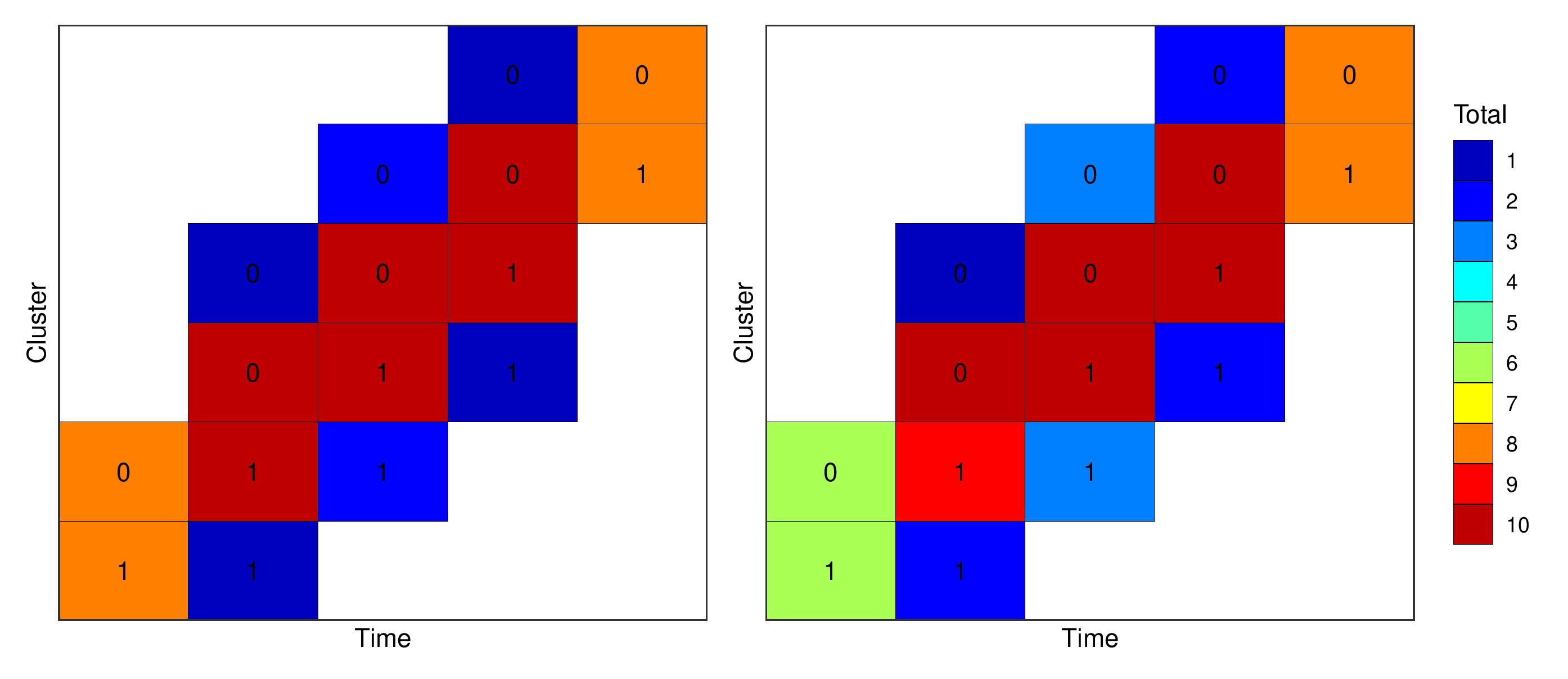}
    \caption{Approximate robust and Bayesian optimal designs for examples D-G. The 0 and 1 labels in the top row indicate the treatment status of the cluster period.}
    \label{fig:optdes3}
\end{figure}

Table \ref{tab:res3} shows the results for the two robust optimal design examples. The results reflect those from all the previous examples: the greedy search performs relatively poorly with variances up to 10\% larger than the best design. Both the local and reverse greedy searches identify the best design in each class. Figure \ref{fig:optdes3} shows the model-robust optimal designs.

\begin{table}
    \centering
    \begin{tabular}{lc|cccccc}
    \toprule
    & Att. & \multicolumn{2}{c}{\textbf{Local}} & \multicolumn{2}{c}{\textbf{Greedy}} & \multicolumn{2}{c}{\textbf{Reverse greedy}} \\
     && \textit{Rel.eff.} & \textit{Time (s)} & \textit{Rel.eff.} & \textit{Time (s)} & \textit{Rel.eff.} & \textit{Time (s)} \\
    \midrule \midrule
       A-D & N/A & 100.0 - 100.3 & 2 - 4 & 102.9 - 109.0 & 0 - 1 &  100.0 & 7  \\
       \midrule
       \multirow{2}{*}{E-H}  & N & 100.0 & 2 - 4 & 104.3 & 0 - 1 & 100.0 & 6\\
         & Y & 100.0 & 2 - 4 & 104.1 & 0 - 1 & 100.0 & 6 \\
       \bottomrule
    \end{tabular}
    \caption{Relative efficiency of the best designs from the three algorithms with and without attenutation for Examples E-H and J with non-Gaussian models.}
    \label{tab:res3}
\end{table}

\section{Conclusion}
\label{sec:conc}
In this article, we have showed that the c-optimal design criterion is a monotone supermodular function for GLMMs using the GLS information matrix. We evaluated the performance of three supermodular function minimisation algorithms to identify c-optimal experimental designs with correlated experimental units. The theoretical upper bound on the relative variance of a design from the local search algorithm is 1.5 and no bound exists for the greedy and reverse greedy algorithms, however, for the examples we considered the performance is significantly better than 1.5 times the best design. The greedy algorithm performed the worst, which was to be expected given that it cannot be executed fully as it cannot start from the empty set. The local search and reverse greedy searches performed comparably in terms of their best designs, although the local search could produce designs with varaince more than 10\% larger than the output of the reverse greedy search. Thus, the local search needs running multiple times to provide a reliable output. The local search also had poorer scaling in terms of computation time than the reverse greedy. Thus, while the reverse greedy search lacks a theoretical guarantee, it would be favoured empirically for the types of study design considered here. 

We showed that the algorithms could also be applied for model robust optimal design identification using a weighted average design criterion. The method uses a prior, specifying the weights to place on each possible model. This specification suggests a way of applying these algorithms for use in Bayesian optimal design. \citet{Chaloner1995} comprehensively review Bayesian experimental design criteria and show the Bayesian c-optimality function to be:
\begin{equation*}
\label{eq:bayes1}
    \int \int (c^T[M_d + V_0]^{-1}c) p(\theta) p(\beta) d\theta d\beta
\end{equation*}
where $V_0$ is the prior covariance of the $\beta$ parameters and $p(\theta)$ and $p(\beta)$ the prior density function for the covariance and linear predictor parameters, respectively. One can approximate the integral above using a Riemann sum, which would discretise the parameter space and provide a set of weights to place on each model. Recent advances have generated general algorithms for Bayesian optimal design problems with non-linear models, in particular \citet{Overstall2017}. Further research is needed to determine whether an approximation using the simple algorithms in this paper provide a viable or useful alternative to more advanced approaches.

We cannot guarantee that the optimal design was included in the output of any of the algorithms. However, our comparison with other multiplicative methods provides some reassurance. For designs with correlation within but not between experimental units, deriving weights using multiplicative methods for each experimental unit provides one method of approximating an optimal design \citep{Holland-Letz2011,Sagnol2011}. Combinatorial approaches produced the same or better designs in the examples we considered. 
 
Optimal designs may sometimes be impractical or difficult to implement. The design in the right panel of Figure \ref{fig:optdes} is highly unlikely to ever be implemented. However, being able to identify approximately optimal designs provides a benchmark against which to justify proposed experiments. Many types of study that can be described by GLMMs, such as cluster randomised trials or spatio-temporal sampling across an area, can be significant and expensive undertakings. The combinatorial algorithms provide a means of identifying near-optimal or optimal designs to support their planning. Many design problems are not inherently discrete; however, we can discretise the design space by specifying a set of design points \citep{Yang2013}. Thus, the methods evaluated in this article provide a useful set of tools to support study design.

\bigskip
\begin{center}
{\large\bf SUPPLEMENTARY MATERIAL}
\end{center}
\section{Rank-1 down/up dating to remove/add observations}
\subsection{Removing an observation}
For a design $d$ with $m$ observations with inverse covariance matrix $\Sigma^{-1}_d$ we can obtain the inverse of the covariance matrix of the design with one observation removed $d' = d / \{i\}$, $\Sigma^{-1}_{d'}$ as follows. Without loss of generality we assume that the observation to be removed is the last row/column of $\Sigma^{-1}_d$. We can write $\Sigma^{-1}_d$ as 
\begin{equation}
    \Sigma^{-1}_d = \begin{pmatrix}
     C & d \\
     d^T & e \\
    \end{pmatrix}
\end{equation}
where $C$ is the $(m-1) \times (m-1)$ principal submatrix of $B$, $d$ is a column vector of length $(m-1)$ and $e$ is a scalar. Then,
\begin{equation}
     G = \Sigma^{-1}_{d / \{i\}} = C - dd^T/e
\end{equation}

\subsection{Adding an observation}
For a design $d$ with $m$ observations with inverse covariance matrix $\Sigma^{-1}_d$, we aim now to obtain the inverse covariance matrix of the design $d' = d \cup \{i'\}$. Recall that $Z$ is a $R \times Q$ design effect matrix with each row corresponding to a possible observation. We want to generate $H^{-1} = \Sigma_{d'}^{-1}$. Note that:
\begin{equation}
    H = \Sigma_{d'} = \begin{pmatrix}
    G^{-1} & f \\
    f^T & h \\
    \end{pmatrix}
\end{equation}
where $f = Z_{i \in d}DZ_{i'}$ is the column vector corresponding to the elements of $\Sigma = W^{-1} + ZDZ^T$ with rows in the current design and column corresponding to $i'$, and $h$ is the scalar $W^{-1}_{i',i'} + Z_{i'}DZ_{i'}^T$. Also now define:
\begin{equation}
    H^* = \begin{pmatrix}
    \Sigma_d & 0 \\
    0 & h 
    \end{pmatrix}
\end{equation}
so that 
\begin{equation}
    H^{* -1} = \begin{pmatrix}
    \Sigma^{-1}_d & 0 \\
    0 & 1/h
    \end{pmatrix}
\end{equation}
and 
\begin{equation}
    H^{**} = \begin{pmatrix}
    \Sigma_d & f \\
    0 & h 
    \end{pmatrix}
\end{equation}
and $u = (f^T, 0)^T$ and $v=(0,...,0,1)^T$, both of which are length $m$ column vectors. So we can get $H^{**}$ from $H^*$ using a rank-1 update as $H^{**} = H^* + uv^T$ and similarly $H = H^{**} + vu^T$. Using the Sherman-Morison formula:
\begin{equation}
    H^{**-1} = H^{* -1} - \frac{H^{* -1}uv^TH^{* -1}}{1+v^TH^{* -1}u}
\end{equation}
and 
\begin{equation}
    H^{-1} = H^{** -1} - \frac{H^{** -1}vu^TH^{** -1}}{1+u^TH^{**-1}v}
\end{equation}
So we have calculated the updated inverse with only matrix-vector multiplication, which is $O(n^2)$.

\begin{description}

\item[Title:] Brief description. (file type)

\item[R-package for  MYNEW routine:] R-package ÒMYNEWÓ containing code to perform the diagnostic methods described in the article. The package also contains all datasets used as examples in the article. (GNU zipped tar file)

\item[HIV data set:] Data set used in the illustration of MYNEW method in Section~ 3.2. (.txt file)

\end{description}

\bibliographystyle{agsm}

\bibliography{opt1}
\end{document}